\documentclass[pdflatex,sn-mathphys-num]{sn-jnl}

\usepackage{graphicx}%
\usepackage{multirow}%
\usepackage{amsmath,amssymb,amsfonts}%
\usepackage{amsthm}%
\usepackage{mathrsfs}%
\usepackage[title]{appendix}%
\usepackage{xcolor}%
\usepackage{textcomp}%
\usepackage{manyfoot}%
\usepackage{booktabs}%
\usepackage{algorithm}%
\usepackage{algorithmicx}%
\usepackage{algpseudocode}%
\usepackage{listings}%

\theoremstyle{thmstyleone}%
\theoremstyle{thmstyletwo}%
\theoremstyle{thmstylethree}%

\begin{document}

\title[Article Title]{Adaptive and accuracy-aware multiple data assimilation in a three step framework}

\author*[1]{\fnm{Kyle} \sur{Ivey}}\email{krivey@ucsd.edu}

\author*[1]{\fnm{Matthias} \sur{Morzfeld}}\email{mmorzfeld@ucsd.edu}

\author[2]{\fnm{Chaoyi} \sur{Wang}}

\author[2]{\fnm{Christina} \sur{Morency}}

\author[2]{\fnm{Christopher S.} \sur{Sherman}}

\author[1]{\fnm{Robert} \sur{Mellors}}

\author[2]{\fnm{Joshua A.} \sur{White}}\email{white230@llnl.gov}

\affil*[1]{\orgdiv{Scripps Institution of Oceanography}, \orgname{University of California, San Diego}, \orgaddress{\street{9500 Gilman Drive}, \city{La Jolla}, \postcode{92093}, \state{CA}, \country{USA}}}

\affil[2]{\orgdiv{Computational Geosciences Group}, \orgname{Lawrence Livermore National Laboratory}, \orgaddress{\street{P.O. Box 808, L-286}, \city{Livermore}, \postcode{94551}, \state{CA}, \country{USA}}}

\abstract{The ensemble smoother with multiple data assimilation (ES-MDA) is an algorithmic framework for the ensemble-based solution of inverse problems in reservoir engineering (and beyond). ES-MDA gradually transitions a prior ensemble to a posterior ensemble. The details of how this transition, or ``multiple data assimilation,'' is implemented defines the accuracy and computational cost of ES-MDA. We show that many popular, adaptive variants of ES-MDA can be understood within a simple three-step framework: inflation proposal, pre-analysis revision, and post-analysis revision. The three steps interact to resolve a trade-off between accuracy (many assimilations with small updates) and efficiency (few assimilations with large updates), inherent to ES-MDA. We then present a new adaptive and accuracy-aware method, ES-MDA-A2, that combines large updates with a ``catch-up'' mechanism that decreases the update size allowing for additional assimilations if the accuracy is low. ES-MDA-A2 requires only two inputs: a targeted accuracy and a maximum number of data assimilations. We test existing and new ES-MDA variants in systematic numerical experiments with a toy model, two electromagnetic inversions with field data, and a subsurface flow reservoir simulation. We find that ES-MDA-A2 resolves the accuracy-efficiency trade-off differently from existing methods, leading to accurate inversions at a reasonable computational cost in all experiments.}



\maketitle

\section{Introduction}

Subsurface property characterization is fundamental to a wide range of geophysical and engineering applications, including hydrocarbon recovery, geothermal energy extraction, and groundwater \citep{essa2025enhanced,de2025ensemble}. 
The underlying inverse problems are difficult and computationally expensive to solve because the forward models are nonlinear and high-dimensional, while observation data are sparse and noisy.
Moreover, robust decision-making requires a rigorous uncertainty quantification (UQ), which is computationally even more expensive than a single solution of an inverse problem.

Ensemble Kalman filters (EnKF) were originally developed in the context of physical oceanography \citep{evensen1994sequential} but have since been widely adopted, e.g., in reservoir engineering (see \cite{emerick2025ensemble}), in climate science \cite{CliMA24}, weather forecasting \citep{HM1998, HEtAl2014}. 
In reservoir engineering, the ensemble smoother (ES, \cite{evensen2000ensemble}) and the ensemble smoother with multiple data assimilation (ES-MDA, \cite{emerick2013ensemble}), are the most popular, workhorse methods because smoothers avoid the computational overhead of repeated simulator restarts by assimilating all data simultaneously.

Mathematically, ES-MDA can be viewed as a sequence of stochastic updates that move the ensemble from a prior distribution toward a Bayesian posterior distribution. This transition is governed by a sequence of \emph{inflation} parameters, which control the step size of each update. When selecting inflation parameters, one faces a fundamental trade-off between computational cost and accuracy.
Small updates are preferable in terms of accuracy, but lead to slow convergence, a large number of iterations and, therefore, a large computational cost.
If the updates are too large, however, the ES-MDA may diverge or lead to nonphysical and inaccurate results.
Current ES-MDA approaches, reviewed in detail in Section~\ref{sec:Background} below, address this trade-off in different ways.
Some use heuristics to define constant or geometrically decaying inflation parameters \citep{rafiee2017theoretical}, others adaptively determine the inflation parameters schedule on the fly \citep{le2016adaptive,iglesias2021adaptive,emerick2016analysis}.

In this work, we provide a unified perspective on ensemble-based smoothing and demonstrate that all ES-MDA variants can be decomposed into three essential steps: (i) inflation proposal, (ii) pre-analysis revision, and (iii) post-analysis revision.
This decomposition clarifies the relationships between existing ES-MDA methods and further elucidates the ``design space'' for new ES-MDA methods and the modification of existing techniques. For example, many adaptive ES-MDA methods enforce monotonically decreasing inflation parameters, which can result in premature termination (poor accuracy) or excessive iteration (wasted compute). While it is intuitive and sensible to make smaller updates early in the ES-MDA iteration, followed by larger updates in later steps, there is an opportunity to consider more general and more flexible approaches.
We consider one such approach which we call ``accuracy aware'' ES-MDA (ES-MDA-A2), in detail.
The two key features of ES-MDA-A2 are that the inflation parameters are chosen based on the current accuracy of the ensemble and that the only user tuning parameters of ES-MDA-A2 are a desired target accuracy and a maximum number of iterations.

The rest of this paper is organized as follows.
In Section~\ref{sec:Background}, we review background materials of ES and variants of ES-MDA. 
In Section~\ref{sec:UnifyingAES}, we present a unifying framework for ES-MDA methods that consists of three essential steps. 
A description of the new, accuracy aware variant of ES-MDA is provided in Section~\ref{sec:ESMDAA2} and numerical experiments are performed in Section~\ref{sec:experiments}.
The numerical examples range from a simple toy model, to inversion of 1D electromagnetic field data and 2D reservoir models.
We conclude the paper in Section~\ref{sec:conclusions} and discuss the cost-accuracy trade-off in adaptive ES-MDA.

\section{Background} \label{sec:Background}
An inverse problem can be formulated as follows: Given a set of noisy data observations $\mathbf{d}_\text{obs}\in\mathbb{R}^{N_d}$ and a forward model $\mathcal{F}(\cdot)$ that maps model parameters $\mathbf{m}\in\mathbb{R}^{N_m}$ to data, find optimal model parameters $\mathbf{m}^\text{opt}$, so that the forward model matches the data,
i.e., $\mathcal{F}(\mathbf{m}^{\text{opt}})\approx \mathbf{d}_\text{obs}$.
The optimal model parameters are obtained by minimizing the objective function
\begin{equation}\label{eq:costfunc}
    \mathcal{O}(\mathbf{m}) = \frac{1}{2} (\mathcal{F}(\mathbf{m})-\mathbf{d}_\text{obs})^\top C_\text{D}^{-1}(\mathcal{F}(\mathbf{m})-\mathbf{d}_\text{obs}) + \frac{1}{2} (\mathbf{m} - \mathbf{m}_{\text{pr}})^\top C_{\text{M}}^{-1}(\mathbf{m} - \mathbf{m}_{\text{pr}});
\end{equation}
here, $C_\text{D}$ is a $N_d\times N_d$ symmetric positive definite (SPD) matrix that describes errors in the data observations,
$\mathbf{m}_\text{pr}$ is a ``prior'' model, and $C_\text{M}$ is a $N_m\times N_m$ SPD matrix that describes uncertainties in the prior model;
throughout, we use superscript $\top$ to denote a transpose.

If we assume that errors in the data observations are Gaussian with mean zero and covariance matrix $C_\text{D}$ and that the prior is Gaussian with mean $\mathbf{m}_\text{pr}$ and prior covariance $C_\text{M}$, 
then the inverse problem defined by the objective function~\eqref{eq:costfunc} has a Bayesian interpretation.
Specifically, the posterior distribution over the model parameters conditioned on the data observations can be written as
\begin{equation}
\label{eq:Posterior}
    p(\mathbf{m}|\mathbf{d}_\text{obs}) \propto p(\mathbf{m})p(\mathbf{d}_\text{obs}|\mathcal{F}(\mathbf{m})) \propto \exp\left(-\mathcal{O}(\mathbf{m})\right).
\end{equation}
The rest of this paper focuses on numerical methods for solving the Bayesian inverse problem.

In the context of reservoir engineering, the forward model $\mathcal{F}(\cdot)$ is often a reservoir simulator that encapsulates the physics describing injection or extraction of fluid within the subsurface and providing numerical predictions of well-based or geophysics-based observations.
The adjoints (gradients) of reservoir simulators are usually unavailable, so that derivative-free ensemble methods have become the workhorse algorithms for the solution of Bayesian inverse problems in reservoir engineering.
An ensemble method represents the posterior distribution~\eqref{eq:Posterior} by an \emph{ensemble}, i.e., a collection of $N_e$ models $\{\mathbf{m}^j\}$, $j=1,\dots,N_e$, which are (approximate) draws from the posterior distribution.

\subsection{Ensemble smoother}

The ensemble smoother (ES) \citep{evensen2000ensemble} assimilates data observations collected during a time interval $[0,T]$, where $T$ could be years or decades (the lifespan of a reservoir). The terminology of a \emph{smoother} is used here because information from data collected at a certain time $t\in[0,T]$ propagates forward and backward in time, influencing the model trajectory over the entire time interval. 
The forward and backward propagation of information from data leads to ``smoother'' trajectories than when data observations are assimilated sequentially in time (filtering).

To set up the ES, we define the predicted data associated with ensemble member $j$ by
\begin{equation}\label{eq:fwdmodel}
    \mathbf{d}^{(j)} = \mathcal{F}(\mathbf{m}^{(j)}),
\end{equation}
and we define the ensemble averages
\begin{equation}
    \overline{\mathbf{m}} = \frac{1}{N_e}\sum_{j=1}^{N_e}\mathbf{m}^{(j)},\quad
    \overline{\mathbf{d}} = \frac{1}{N_e}\sum_{j=1}^{N_e}\mathbf{d}^{(j)},
\end{equation}
for the ensembles of models and predicted data.
With these definitions, we can define the model-data cross-covariance matrix $C_\text{MD}\in\mathbb{R}^{N_m\times N_d}$  by
\begin{equation}
    C_{\text{MD}} = \frac{1}{N_e-1}\sum_{j=1}^{N_e}
    \left(\mathbf{m}^{(j)}-\overline{\mathbf{m}}\right)
    \left(\mathbf{d}^{(j)} - \overline{\mathbf{d}} \right)^\top,
\end{equation}
and the data-data auto-covariance matrix $C_\text{DD}\in\mathbb{R}^{N_d\times N_d}$ by 
\begin{equation}
        C_{\text{DD}} = \frac{1}{N_e-1}\sum_{j=1}^{N_e}
        \left(\mathbf{d}^{(j)} - \overline{\mathbf{d}}\right)
        \left(\mathbf{d}^{(j)} - \overline{\mathbf{d}}\right)^\top.
\end{equation}
The ES update equation for ensemble member $j$ is 
\begin{equation}\label{eq:ES}
    \hat{\mathbf{m}}^{(j)} = \mathbf{m}^{(j)} + C_\text{MD}(C_\text{DD} + C_\text{D})^{-1}(\mathbf{d}_\text{obs}^{(j)} + \boldsymbol{\eta}^{(j)} - \mathcal{F}(\mathbf{m}^{(j)})),
\end{equation}
where $\boldsymbol{\eta}\sim\mathcal{N}(\mathbf{0},C_\text{D})$ is a Gaussian random variable with mean zero and covariance matrix $C_\text{D}$.
If the forward model $\mathcal{F}(\cdot)$ is linear, then the posterior distribution~\eqref{eq:Posterior} is also Gaussian and the ES draws samples from that distribution, provided the ensemble size $N_e$ is sufficiently large.

\subsection{Ensemble smoother with multiple data assimilation}
Recall that the posterior distribution~\eqref{eq:Posterior} is the product of a prior and a likelihood. 
The likelihood thus ``transitions'' the prior distribution to the posterior distribution.
In nonlinear\slash non-Gaussian problems, it can be advantageous to gradually introduce the data observations by re-writing the posterior distribution as
\begin{equation}
\label{eq:TemperedLikelihood}
    p(\mathbf{m}|\mathbf{d}_\text{obs}) \propto p(\mathbf{m})p(\mathbf{d}_\text{obs}|\mathcal{F}(\mathbf{m}))^{1/\alpha_1}
    p(\mathbf{d}_\text{obs}|\mathcal{F}(\mathbf{m}))^{1/\alpha_2}\cdots p(\mathbf{d}_\text{obs}|\mathcal{F}(\mathbf{m}))^{1/\alpha_{N_a}},
\end{equation}
where the scalars $\alpha_n>1$ are chosen such that 
\begin{equation}\label{eq:alphacriterion}
    \sum_{n=1}^{N_a} \frac{1}{\alpha_n} = 1.
\end{equation}
It is convenient to track this constraint using the cumulative inverse inflation parameter
\begin{equation}\label{eq:inverse_inflation}
    \beta_n = \sum_{i=1}^n \frac{1}{\alpha_n}, \quad n>0,
\end{equation}
where $\beta_0=0$ and ES-MDA should stop iterating once $\beta_n=1$.
Using~\eqref{eq:TemperedLikelihood} along with a ``sequential'' data assimilation now allows for first transitioning from the prior, to an intermediate posterior $p(\mathbf{m})p(\mathbf{d}_\text{obs}\vert \mathbf{m})^{1/\alpha_1}$, which turns into a prior for the next posterior $p(\mathbf{m})p(\mathbf{d}_\text{obs}\vert \mathbf{m})^{1/\alpha_1}p(\mathbf{d}_\text{obs}\vert \mathbf{m})^{1/\alpha_2}$, and so on.
In this way, one can slowly transition form the prior to the posterior distribution by performing $N_a$ data assimilation steps.
Applying the ES to each intermediate posterior distribution results in the algorithm known as ensemble smoother with multiple data assimilation (ES-MDA) of \cite{emerick2013ensemble}.
Specifically, the ES update is applied iteratively in $N_a$ steps
\begin{equation} \label{eq:ESMDA}
    \mathbf{m}_{n+1}^{(j)} = \mathbf{m}_{n}^{(j)} + C_{\text{MD}}(C_{\text{DD}} + \alpha_{n}C_\text{D})^{-1}\left(\mathbf{d}_{\text{obs}}^{(j)} + \sqrt{\alpha_n}\boldsymbol{\eta}_n^{(j)} - \mathcal{F}(\mathbf{m}_{n}^{(j)})\right),
\end{equation}
where $n=1,\dots,N_a$ is the iteration index. 
If we set $N_a=1$, ES-MDA is just the ES discussed above (see also \cite{emerick2012history}).

ES-MDA and the implied gradual transition from prior to posterior through incremental updates is beneficial in nonlinear models for two reasons: (i) inflating the data covariance $C_D$ by $\alpha_n$ reduces the condition number of the  matrix $C_\text{DD}+\alpha_n C_\text{D}$ in the update equation~\eqref{eq:ESMDA}; (ii) a gradual transition from prior to posterior helps to alleviate the nonlinearity of the forward model, which enters through the likelihood. 

\subsection{Variations of ES-MDA}
Nonlinear problems benefit from iterating and gradually transitioning from prior to posterior, hence the choice of the \emph{inflation }parameters $\alpha_n$ is critical in ES-MDA.
While in theory \emph{any} choice that satisfies~\eqref{eq:alphacriterion} is valid, the specifics of how the inflation parameters are chosen defines the efficiency and accuracy of ES-MDA.

\subsubsection{ES-MDA with a fixed number of assimilations}
Possibly the simplest choice for the inflation parameters is to choose a constant $\alpha_n = N_a$ \citep{emerick2013ensemble}.
Alternatively, one can consider geometrically decreasing inflation parameters 
\begin{equation}\label{eq:alphaGEO}
    \alpha_{n+1} = \gamma^n \alpha_1,
\end{equation}
where $\gamma_1,\dots,\gamma_{N_a-1}$ are chosen such that $\alpha_1,\dots,\alpha_{N_a}$ are a decreasing sequence \citep{rafiee2017theoretical};
specifically, one can determine $\gamma^n$ by solving
\begin{equation}
    f(\gamma) = \sum_{n=1}^{N_a}\frac{1}{\gamma^{n-1}\alpha_1}-1 = 0
\end{equation}
for a given $\alpha_1$ and a given number of assimilations $N_a$.
Geometrically decaying inflation parameters make small updates (large $\alpha_n$) early on in the iteration and larger updates (small $\alpha_n$) later during the iteration, which is perhaps intuitive \citep{rafiee2017theoretical,emerick2016analysis, iglesias2015iterative}.
\cite{evensen2018analysis} found that the benefit of geometrically decaying inflation parameters is more pronounced when fewer assimilations are performed (small $N_a$). 

With either constant or geometrically decaying inflation parameters, the number of iterations $N_a$ needs to be tuned, which can waste computations.
If an initial choice for $N_a$ (typically $N_a=4$) does not lead to a satisfactory accuracy (e.g., large data mismatch after ES-MDA), then another ES-MDA has to be performed with a larger number of assimilations, but the results of the initial ES-MDA cannot be reused.

\subsubsection{Adaptive ES-MDA: Determining the number of assimilations during the iteration}\label{sec:adaptive_esmda}
Several ES-MDA methods that adaptively determine $\alpha_n$ \emph{without} specifying the number of assimilations have been proposed over the years.
Examples include ES-MDA adapt \citep{emerick2016analysis}, ES-MDA-RLM, ES-MDA-RS \citep{le2016adaptive} and EKI-DMC \cite{iglesias2021adaptive}.

We will show below that \emph{all} ES-MDA methods (adaptive or not) can be formulated within a unified framework that consists of three steps: 
\begin{enumerate}[label=(\roman*)]
    \item propose an inflation parameter; 
    \item check and revise the proposal prior to assimilation; 
    \item check and revise the proposal after assimilation.
\end{enumerate}
The various adaptive ES-MDA techniques only differ in how these steps are executed. 
ES-MDA techniques with a fixed number of assimilations simply skip the checks before and after the assimilation and always accept the (constant or geometrically decaying) inflation parameters.

There is also related work on the iterative ensemble smoother (iES) \citep{ma2019robust},
which determines the inflation parameter $\alpha_n$ during an iteration based on theory borrowed from step-size control strategies for Levenberg-Marquardt trust region optimization.
This approach relies on a monotonic decrease of the objective function~\eqref{eq:costfunc}, requiring line searches and repeated ensemble evaluations, making iES often more costly than ES-MDA. Additionally, iES methods have been shown to require many iterations and they sample a similar, but different, approximate posterior distribution than ES-MDA.
 
\subsection{Metrics of data misfit}
\label{sec:Cost}
Some adaptive ES-MDA methods determine inflation parameters based on the mismatch of predicted and observed data,
but use different metrics of data mismatch, which we briefly review here. 
The mean-square-error (MSE) of an ensemble member ($j$) at iteration ($n$) and the corresponding ensemble averaged MSE are 
\begin{equation}
        \mathcal{M}_n^{(j)} = \frac{1}{N_d}\left\|C_\text{D}^{-1/2}\left(\mathbf{d}_n^{(j)}-\mathbf{d}_{obs}\right)\right\|^2,
        \quad
        \overline{\mathcal{M}}_n = \frac{1}{N_e}\sum_{j=1}^{N_e}\mathcal{M}_n^{(j)},
\end{equation}
where the vertical bars denote the two-norm ($\| x \| = \sqrt{x^\top x}$).
MSE is popular in the data assimilation and computer science communities, but its square root, the root mean square error (RMSE) is more popular in geophysics \citep[see, e.g.,][]{constable1987occam}.
In our notation, the RMSE of ensemble member $j$ at iteration $n$ and its ensemble average are
\begin{equation}
    \mathcal{R}_n^{(j)} =\sqrt{\mathcal{M}_n^{(j)}} = \frac{1}{\sqrt{N_d}}\left\| C_\text{D}^{-1/2}\left(\mathbf{d}_n^{(j)}-\mathbf{d}_{obs}\right)\right\|,
    \quad
    \overline{\mathcal{R}}_n = \frac{1}{N_e}\sum_{j=1}^{N_e}\mathcal{R}_n^{(j)}.
\end{equation}
One can also consider the RMSE associated with the ensemble average $\overline{\mathbf{m}}$, defined by
\begin{equation}
    \label{eq:RMSEofAvg}
    \hat{\mathcal{R}} =\frac{1}{\sqrt{N_d}} \left\| C_{\text{D}}^{-1/2}\left( \mathcal{F}(\overline{\mathbf{m}})-\mathbf{d}_{\text{obs}}\right) \right\|.
\end{equation}
The adaptive ES-MDA method (EKI-DMC, \cite{iglesias2021adaptive}) computes the ``un-normalized MSE'' and its ensemble average at iteration $n$, defined by
\begin{equation}
\label{eq:Phi}
    \Phi_n^{(j)} = 
    N_d\mathcal{M}_n^{(j)} = 
    \left\|C_\text{D}^{-1/2}\left(\mathbf{d}_n^{(j)}-\mathbf{d}_{obs}\right)\right\|^2,\quad
    \overline{\Phi}_n = \frac{1}{N_e}\sum_{j=1}^{N_e}\Phi_n^{(j)}.
\end{equation}
We will show below that the choice of misfit measure influences the efficiency and accuracy of adaptive ES-MDA methods. 

\section{Unifying ES-MDA in three essential steps}
\label{sec:UnifyingAES}
We now formalize the three steps identified in Section~\ref{sec:adaptive_esmda} as the general procedure summarized in Algorithm~\ref{alg:unif}.

\subsection{The three essential steps of ES-MDA}
First, all ES-MDA techniques perform a simulation with each ensemble member using the forward model, and subsequently perform an analysis via equation~\eqref{eq:ESMDA}.
In between simulation and analysis, ES-MDA performs two steps.

\vspace{2mm}
{\bf{Step 1: Inflation proposal}}.
To perform an analysis, an ES-MDA requires an inflation parameter $\alpha_n$.
Simple strategies may propose a constant or geometrically decaying $\alpha_n$, but adaptive methods (ES-MDA adapt, ES-MDA-RLM and ES-MDA-RS, EKI-DMC) propose $\alpha_n$ based on data misfit.

\vspace{2mm}
{\bf{Step 2: Pre-analysis inflation revision}}.
The proposed inflation parameter should be revised if it violates the condition for appropriate spread in equation~\eqref{eq:alphacriterion} (equivalently, $\beta_n>1$).
The common strategy here is to choose an $\alpha_n$ such that $\beta_{n-1} + 1/\alpha_n=1$, ensuring that the iteration terminates after the next analysis.
Non-adaptive ES-MDA (constant or geometrically decaying $\alpha_n$) do not need to revise the inflation parameter proposal.

\vspace{2mm}
{\bf{Analysis}}.
Once an appropriate inflation parameter is set, ES-MDA performs an analysis using equation~\eqref{eq:ESMDA}. 

\vspace{2mm}
{\bf{Step 3: Post-analysis inflation revision}}.
Several ES-MDA techniques, e.g., ES-MDA-RS or ES-MDA-RLM, assess the quality of the analysis ensemble and may revise the inflation parameter if the analysis ensemble changed too much, or not enough, or if no significant change in a data misfit metric is detected.
Revising the inflation parameter after the analysis is computationally more costly than a pre-analysis revision, because the analysis has to be re-computed after the inflation parameter was modified.

\vspace{2mm}
All ES-MDA methods stop iterating once condition~\eqref{eq:alphacriterion} is satisfied (alternatively, once $\beta_n=1$).
We summarize the essential steps of ES-MDA in Algorithm~\ref{alg:unif}.

\begin{algorithm}[tb]
\caption{Unified ES-MDA}\label{alg:unif}
\begin{algorithmic}[1]
\Require Initial ensemble $\{\mathbf{m}_0^j\}_{j=1}^{N_e}$, maximum iterations $N_{\max}$
\Ensure Updated ensemble
\State Initialize $n \leftarrow 0$, $\alpha_0 \leftarrow 0$, $\beta_0 \leftarrow 0$
\While{$n \le N_{\max}$}
    \State $n \leftarrow n+1$
    \State \textbf{Simulation}
    \State Run the forward model for each ensemble member $j$ 
    \State 
    \State \textbf{Step 1: Inflation proposal}
    \State Propose $\alpha_n'$ and set $\beta_n' = \beta_{n-1} + 1/\alpha_n'$
    \State

    \State \textbf{Step 2: Pre-analysis inflation revision}
    \If{$\beta_n'>1$ (or, e.g., $\alpha_n'>\alpha_\text{max}$)}
        \State Revise $\alpha_n'$ and set $\beta_n' = \beta_{n-1} + 1/\alpha_n'$  
    \EndIf

    \State
    \State \textbf{Analysis}
    \State Update ensemble using the analysis equation~\eqref{eq:ESMDA} with $\alpha_n'$

    \State
    \State \textbf{Step 3: Post-analysis inflation revision}
    \If{analysis rejected}
        \State Revise $\alpha_n'$ and set $\beta_n' = \beta_{n-1} + 1/\alpha_n'$  
        \State \textbf{go to} Analysis
    \EndIf

    \State
    \State Set $\alpha_n \leftarrow \alpha_n'$, $\beta_n \leftarrow \beta_{n}'$
    \If{$\beta_n = 1$}
        \State \textbf{break}
    \EndIf
\EndWhile
\end{algorithmic}
\end{algorithm}

\subsection{Examples}
We now explain how various ES-MDA techniques implement the three steps of ES-MDA described above, since this implementation defines each methods accuracy, efficiency, and tuning requirements.

\subsubsection{ES-MDA with constant or geometrically decaying inflation parameters (Step 1)}
If we specify a number of data assimilations $N_a$ and pre-compute a sequence of $N_a$ inflation parameters, then we can skip the revision steps (Steps~2 and~3). 
Such a simple procedure requires that the number of assimilations $N_a$ be tuned.
Note that the revision steps are typically computationally less costly than forward model evaluations, so that the computational gains resulting from a simple inflation scheme are minimal.
Simple inflation schemes in ES-MDA are therefore analogous to performing a gradient-based optimization without step-size control or convergence checks, which is unusual in optimization.

\subsubsection{ES-MDA with pre-analysis inflation revision (Steps~1 and~2): ES-MDA adapt and EKI-DMC}\label{sec:322}
ES-MDA adapt \cite{emerick2016analysis} and EKI-DMC \cite{iglesias2021adaptive} propose an inflation parameter based on a data misfit metric (Step~1) and revise the proposal (Step 2) $\alpha_n$ as $\beta_{n-1} + 1/\alpha_n=1$ if the condition~\eqref{eq:alphacriterion} is violated to terminate after the next analysis and ensure that the final analysis ensemble has an appropriate spread. ES-MDA adapt proposes an inflation parameter using the MSE to measure data misfit (step~1):
\begin{equation}\label{eq:alphaAdapt}
    \alpha_n = a\cdot \overline{\mathcal{M}}_n,
\end{equation}
where $a>0$ is a tuneable parameter (and $a=0.25$ may be a good choice, see below).
The proposed inflation parameter is revised (step~2) if it exceeds a maximum value, in which case the proposal is revised to be the maximum allowed value $\alpha_n=\alpha_{\max}$.
In terms of tuning, ES-MDA adapt requires determining the factor $a$ (which is often set to $a=0.25$) and the maximum inflation parameter $\alpha_{\max}$.

EKI-DMC operates with the same Step 2 revision as ES-MDA adapt, but the proposal mechanism for $\alpha_n$ (Step 1) is based on the unscaled data misfit metric $\Phi_n$ in~\eqref{eq:Phi}.
Specifically, the proposed inflation parameter is
\begin{equation}\label{eq:alphaEKIDMC}
    \alpha_{n}^{-1} = \text{min}\left\{\text{max}\left\{{\frac{N_d}{\overline{\Phi}_n},
    \sqrt{\frac{N_d}{2 \sigma^2(\Phi_n) }}}\right\},
    1-\beta_n\right\},
\end{equation}
where 
\begin{equation}
    \sigma^2(\Phi_n)=\frac{1}{N_e-1}\sum_{i=1}^{N_e} \left(\Phi_n^{(j)}-\overline{\Phi}_n\right)^2,
\end{equation}
is the ensemble variance of the data misfit $\Phi_n$.
This proposal is motivated by controlling the Kullback–Leibler divergence between successive ensemble distributions \citep{iglesias2021adaptive}.
EKI-DMC does not have tunable parameters, but in the numerical experiments below we find that EKI-DMC tends to perform many iterations, favoring accuracy over computational efficiency.

\subsubsection{ES-MDA with pre- and post-analysis inflation revision (Steps~1-3): ES-MDA-RS and ES-MDA-RLM}
Two of the ES-MDA techniques we consider perform all three steps of ES-MDA, including a revision of the inflation parameter after the analysis has been performed. 
Both methods, ES-MDA-RS and ES-MDA-RLM \citep{le2016adaptive} first propose an inflation parameter in the same way as ES-MDA adapt, but with $a=0.25$ so that the proposal is
\begin{equation}
    \alpha_n = 0.25\cdot \overline{\mathcal{M}_n}.
\end{equation}
The proposal is revised in Step~2 the same as in Section~\ref{sec:322}.
The difference between ES-MDA adapt and ES-MDA-RS\slash ES-MDA-RLM is step 3. 
Specifically, ES-MDA-RS checks, post-analysis, if the maximum change in any of the model parameters exceeds twice the standard deviation of the model parameters at step $n-1$. 
If the maximum change exceeds this threshold, the analysis is performed again, but with $2\cdot \alpha_n$. In our notation, the post-analysis check of ES-MDA-RLM is as follows.
Check that, for each ensemble member
\begin{equation}
\label{eq:ESMDARLM}
    \begin{split}
    \rho^2\left\|
    C_D^{-1/2}\left(\mathbf{d}_{\text{obs}}^{(j)} + \sqrt{\alpha_n}\boldsymbol{\eta}_n^{(j)} - \mathcal{F}(\mathbf{m}_{n}^{(j)})\right)
    \right\|^2 &< \\
    \alpha_n^2\left\|C_D^{1/2}(C_{\text{DD}} + \alpha_{n}C_\text{D})^{-1}
    (\mathbf{d}_{\text{obs}}^{(j)} + \sqrt{\alpha_n}\boldsymbol{\eta}_n^{(j)} - \mathcal{F}(\mathbf{m}_{n}^{(j)}))\right\|^2,
    \end{split}
\end{equation}
where $\rho=0.5$ (but other choices are also possible).
If this condition is violated for any ensemble member, then double $\alpha_n$ (for all ensemble members) and re-compute until no violation occurs.
\cite{le2016adaptive} also propose an alternative, which is a slight variation of the above step~3. Procedures of this kind are referred to as ``step-size cuts'' and in practice must be bounded by defining a maximum number of step-size cuts to avoid $\alpha_n\rightarrow\infty$.

ES-MDA-RS and ES-MDA-RLM have no free tuning parameters, but both methods have intrinsically made choices for some tunable parameters ($\alpha_{\max}$ and $a=0.25$ for both and $\rho=0.5$ for ES-MDA-RLM).

\section{Accuracy-aware ES-MDA}
\label{sec:ESMDAA2}
We introduce a new variation of ES-MDA that is inspired by two key insights we gained from comparing the various ES-MDA algorithms and their three essential steps. First, we use RMSE to propose an inflation parameter (Step~1). Since RMSE is smaller than MSE, this choice leads to smaller inflation parameters $\alpha_n$ and hence larger and more aggressive updates compared to ES-MDA adapt, ES-MDA-RS, ES-MDA-RLM or EKI-DMC. Second, we revise the inflation parameter based on data misfit (Steps~2 and~3). Revising the inflation parameter using data misfit is not a new concept in ES-MDA, but it is intuitive -- once the data misfit is small, the inflation should be small so that the iteration can terminate. 
On the other hand, if the data misfit is large, the inflation should increase to enable more iterations.
Since our new algorithm modifies the inflation parameter based on accuracy, we term our implementation ``accuracy-aware ES-MDA'' (ES-MDA-A2).

We now describe the three essential steps of ES-MDA-A2.
In step~1, we propose an inflation parameter proportional to the average RMSE,
\begin{equation}
\label{eq:ESMDAA2Prop}
    \alpha_n = a\cdot \overline{\mathcal{R}}_n,
\end{equation}
where we set $a=1$ by default, but other choices are also possible.
In step~2, we consider two scenarios when the proposed $\alpha_n$ violates the condition~\eqref{eq:alphacriterion}.
\begin{enumerate}
    \item If $\vert \overline{\mathcal{R}}_n-\mathcal{R}_\text{target}\vert \leq \tau_\text{target}$, we revise $\alpha_n$ such that $\beta_{n-1} + 1/\alpha_n=1$, ensuring that the iteration terminates after the next analysis. Thus, if the iteration has reached a target misfit, the algorithm terminates.
    \item If the target RMSE is not reached ($\vert \overline{\mathcal{R}}_n-\mathcal{R}_\text{target}\vert >\tau_\text{target}$), ES-MDA-A2 executes a ``catch-up step'' with an increased $\alpha_n$, to allow for more iterations to reach the target RMSE. The catch-up step is described in detail further further below.
\end{enumerate}
After the analysis, we perform step~3, which is also accuracy aware.
But rather than using the average RMSE, we consider the RMSE of the ensemble average~\eqref{eq:RMSEofAvg} to reduce computational costs -- computing the average of the RMSEs of the ensemble members requires $N_e$ simulations, but computing the RMSE of the ensemble average only requires one additional simulation.
With $\hat{\mathcal{R}}$ computed, step~3 amounts to recomputing the analysis with a revised $\alpha_n = 1/(1-\beta_n)$ if $\|\hat{\mathcal{R}}-\mathcal{R}_\text{target}\|\leq \tau_{\text{target}}$ (current analysis is close to the target accuracy) or if the average RMSE does not decrease much, i.e., if $\vert \overline{\mathcal{R}}_n-\overline{\mathcal{R}}_{n-1}\vert \leq \tau_\mathcal{R}$.

The ``catch-up'' step increases the inflation parameter if the target RMSE has not been reached while condition~\eqref{eq:alphacriterion} is violated.
In the catch-up step, we store the iteration number $n^*$, which first triggered a catch-up step and the corresponding $\alpha^*=1/(1-\beta_{n^*-1})$.
We then increase the inflation parameter such that
\begin{equation}\label{eq:ESMDAA2}
    \alpha_n = 2 \cdot \min\left(\frac{1}{1-\beta_{n-1}} , (N_{\max} - n^*) \alpha^* \right).
\end{equation}
The above equation effectively doubles the inflation $\alpha^{\ast}$ during each catch-up step, until a maximum number of iterations $N_{\max}$ is reached. 
ES-MDA-A2 is summarized in pseudo-code in Algorithm~\ref{alg:AES}.

Note that ES-MDA-A2, by design, has two distinct phases: first, large updates are performed because we propose small inflation parameters based on RMSE (rather than based on MSE, which leads to smaller updates). If the target RMSE can be reached within a few large updates, the iteration terminates once the RMSE is near a target RMSE. If, however, the first few large updates lead to a violation of~\eqref{eq:alphacriterion} without reaching the target RMSE, then the inflation increases until a maximum number of iterations is reached. 
In doing so, ES-MDA-A2 allows for additional iterations necessary to  reach a desired accuracy (target RMSE) when the inflation proposals decay too quickly. Our contribution of increasing the inflation parameter in view of a targeted accuracy is unique to ES-MDA-A2. 

\begin{algorithm}[]
\caption{Accuracy-aware ES-MDA}\label{alg:AES}
\begin{algorithmic}[1]
\Require Initial ensemble $\{\mathbf{m}_0^j\}_{j=1}^{N_e}$, maximum iterations $N_{\max}$, target RMSE $\mathcal{R}_{\text{target}}$, tolerances $\tau_\text{target},\tau_\mathcal{R}$
\Ensure Updated ensemble
\State Initialize $n \leftarrow 0$, $\alpha_0 \leftarrow 0$, $\beta_0 \leftarrow 0$
\While{$n \le N_{\max}$}
    \State $n \leftarrow n+1$
    \State \textbf{Simulation}
    \State Run the forward model for each ensemble member $j$
    \State
    
    \State \textbf{Step 1: Inflation proposal}
    \State Propose $\alpha_n' = \overline{\mathcal{R}}_n$ and set $\beta_n' = \beta_{n-1} + 1/\alpha_n'$
    \State
    \State \textbf{Step 2: Pre-analysis inflation revision}
    \If{$\beta_n' > 1$}
        \If{$\vert \overline{\mathcal{R}}_n-\mathcal{R}_\text{target}\vert \leq \tau_\text{target}$}
            \State Revise $\alpha_n' = \dfrac{1}{1-\beta_{n-1}}$ (terminate after next analysis)
        \Else
            \State Revise $\alpha_n' = 2 \cdot \min\left(\dfrac{1}{1-\beta_{n-1}} , (N_{\max} - n^*) \alpha^* \right)$ (catch-up step)
        \EndIf
    \EndIf
    \State 
    
    \State \textbf{Analysis}
    \State Update ensemble using the analysis equation~\eqref{eq:ESMDA} with $\alpha_n'$

    \State
    \State \textbf{Step 3: Post-analysis inflation revision}
    \State Compute $\hat{\mathcal{R}}_n$ (RMSE of ensemble average)
    \If{$\vert\hat{\mathcal{R}}_n-\mathcal{R}_{\text{target}}\vert\leq \tau_{\text{target}}$ \textbf{or} $\vert\hat{\mathcal{R}}_n-\hat{\mathcal{R}}_{n-1}\vert\leq\tau_{\mathcal{R}}$}
        \State Revise $\alpha_n' = \dfrac{1}{1-\beta_{n-1}}$ and set $\beta_n'=\beta_{n-1}+1/\alpha_n'$
        \State \textbf{go to} Analysis (terminate after next analysis)
    \EndIf

    \State
    \State Set $\alpha_n \leftarrow \alpha_n'$, $\beta_n \leftarrow \beta_{n}'$
    \If{$\beta_n = 1$}
        \State \textbf{break}
    \EndIf
\EndWhile
\end{algorithmic}
\end{algorithm}

ES-MDA-A2 requires that we specify a target RMSE $\mathcal{R}_{\text{target}}$ and a maximum number of iterations (the tolerances $\tau_{\text{target}}$ and $\tau_\mathcal{R}$ can be set to default small values).
By design, both inputs have a clearly interpretable meaning, allowing the user to balance accuracy and computational efficiency. 
If accuracy is very important, we set $\mathcal{R}_{\text{target}}=1$ and pick a large number of maximum iterations (but ES-MDA-A2 may still terminate the iteration quickly if the target RMSE can be reached after a few iterations).
If computations are a concern, then $N_{\max}$ should be a small number and ES-MDA-A2 will attempt to determine a good ensemble given the computational constraints.

In the numerical experiments below, ES-MDA-A2 performs well with the default choice of $a=1$ in the inflation proposal step~\eqref{eq:ESMDAA2Prop}.
It is, however, possible that this choice leads to early updates that are too large to achieve a desired target accuracy. 
In that case, one may consider using $a>1$, which transitions ES-MDA-A2 closer to EKI-DMC and further promotes accuracy over efficiency.
Inverse problems with a strong prior, however, are common and ES-MDA then benefits from large early updates ($a=1$).

In practice, ES-MDA methods are localized to increase the accuracy of covariance estimates or Kalman gain approximations.
One might consider the effect of localization and sub-space approaches on how inflation is decided within adaptive ES-MDA methods. 
Although not explored here, we expect that conditioning provided via localization would reduce the need for large inflation. In turn, this would imply that the small inflation parameters of ES-MDA-A2 work well, supporting the choice of smaller inflation provided from RMSE-based proposals.
Finally, we note that the pre-analysis catch-up mechanism could be incorporated into existing algorithms to propose further adaptive ES-MDA methods. 
We do not pursue this idea in this paper, but future work could investigate the accuracy and cost of existing methods with this added element.

\section{Numerical experiments}
\label{sec:experiments}
We perform a set of systematic numerical experiments to compare and contrast several variants of ES-MDA with the fixed setup outlined in Table~\ref{tab:methods}.
The numerical experiments involve forward models of increasing complexity.
We start with a simple toy problem, then move on to 1D inversions of electromagnetic data and finally present results with 2D compositional flow models.
The experiments illustrate how different ES-MDA schedulers resolve the trade-off between accuracy and computational efficiency in different ways.

\begin{table}[h]
\caption{Experimental setup of ES-MDA variants and related equations. Tolerances are set to default values of $\tau_{\text{target}} = \tau_\mathcal{R} = 0.05$.}\label{tab:methods}%
\begin{tabular}{@{}llll@{}}
\toprule
\textbf{Method} & \textbf{Type} & \textbf{Key Parameters} & \textbf{Reference} \\
\midrule
ES-MDA (4x) & Fixed & $N_a = 4$, $\alpha_n = N_a$ & Eq.~\eqref{eq:ESMDA} \\
ES-MDA (8x) & Fixed & $N_a = 8$, $\alpha_n = N_a$ & Eq.~\eqref{eq:ESMDA} \\
ES-MDA-RS   & Adaptive & $a = 0.25$, max cuts $= 5$ & Eq.~\eqref{eq:alphaAdapt} \\
ES-MDA-RLM  & Adaptive & $a = 0.25$, max cuts $= 5$, $\rho = 0.5$ & Eq.~\eqref{eq:alphaAdapt}; Eq.~\eqref{eq:ESMDARLM} \\
EKI-DMC     & Adaptive & No free parameters & Eq.~\eqref{eq:alphaEKIDMC} \\
ES-MDA-A2   & Adaptive & $\mathcal{R}_{\text{target}} = 1.0$, $N_{\text{max}} = 10$ & Eq.~\eqref{eq:ESMDAA2Prop}; Eq.~\eqref{eq:ESMDAA2} \\
\botrule
\end{tabular}
\end{table}

\subsection{Experiments with a BOD Model}\label{sec:BOD}
We follow \cite{bardsley2014randomize} and consider a computationally inexpensive Biochemical Oxygen Demand (BOD) model. A BOD model describes the biological oxidation of organic matter in a water sample over time, but here we use it as a simple toy problem to compare variants of ES-MDA on a low-dimensional nonlinear problem.
The forward model is 
\begin{equation}
    \mathcal{G}(\boldsymbol{\theta}) = \theta_1(1-\exp(-\theta_2\mathbf{x})),
\end{equation}
where $\boldsymbol{\theta}=(\theta_1,\theta_2)^\top$ are the model parameters we wish to recover via ES-MDA and where $\mathbf{x}=(1,3,5,7,9)^\top$.
In the numerical experiments with synthetic data, the true parameters are $\mathbf{\boldsymbol{\theta}}_{\ast}=(1,0.1)^\top$ and the data observations are 
\begin{equation}\nonumber
    d = (0.076, 0.258, 0.369, 0.492, 0.55)^\top.
\end{equation}
 
For all ES-MDA methods, we generate an initial ensemble by sampling the Gaussian distributions $\theta_1\sim\mathcal{N}(1,1)$ and $\theta_2\sim\mathcal{N}(0.1,0.05)$ and set the ensemble size to $N_e=100$. We repeat each experiment 100 times, because the results depend on the quality of the initial ensemble (which here is chosen at random).
We then compare the computational cost and accuracy of each method.
Accuracy is defined by the average RMSE of the ensemble and the computational cost is approximated by the mean number of iterations.

\begin{figure}[!htb]
 \centering
 \includegraphics[width=\linewidth]{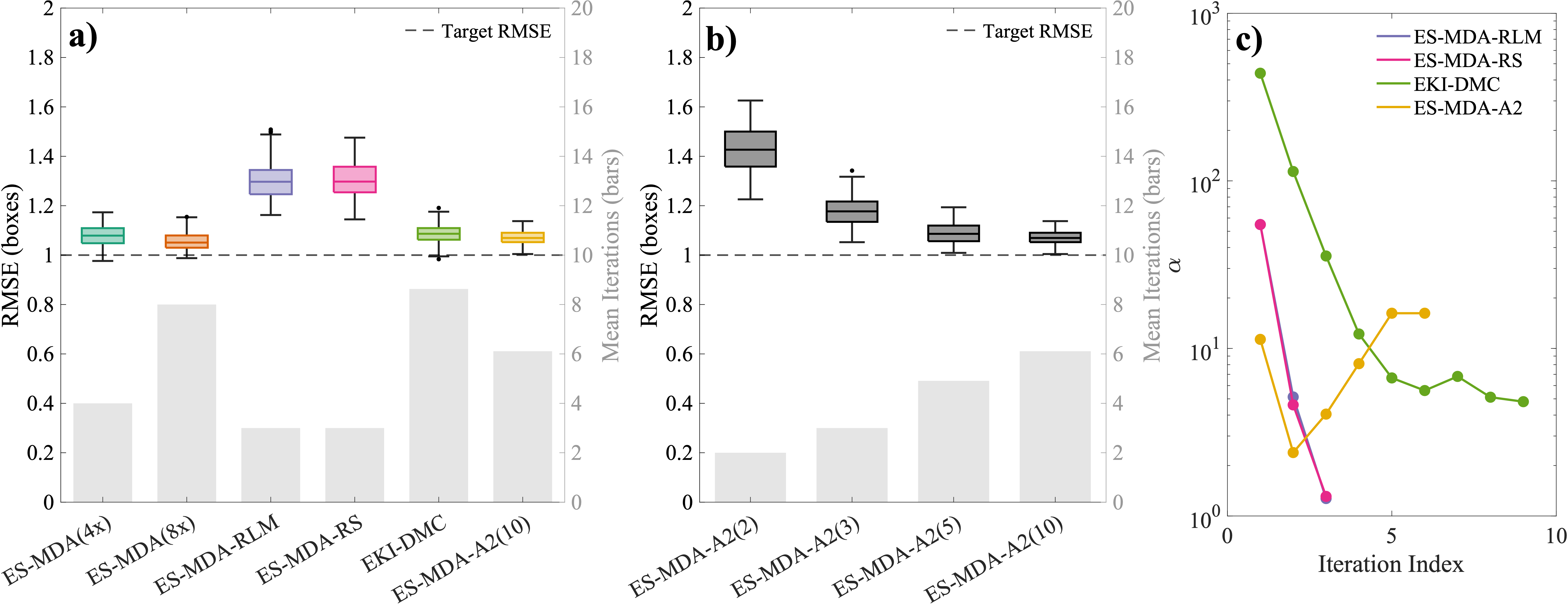}
 \caption{Summary of the numerical experiments with the BOD model.
 (a) Box-plots of RMSE (left y-axis) and bar-charts of the average number of iterations (right y-axis).
 For the box-plots: The box represents the interquartile range (IRQ), whiskers are calculated as $1.5\cdot \text{IQR}$, the median is shown as a solid colored line, and outliers shown as black dots.
 (b) Box-plots and bar-charts as in (a), but for ES-MDA-A2 in different configurations ($N_{\max}={2,3,5,10}$).
 (c) Inflation parameter $\alpha_n$ as a function of the iterations index for the first of our 100 experiments.
 }
 \label{fig:BODresults}
\end{figure}

Figure \ref{fig:BODresults}(a) summarizes the results of our experiments with the BOD model.
The figure combines accuracy and computational cost and shows box-plots of the RMSE and bar-charts of the average number of iterations (panel (a)). 
Also shown are the inflation parameter $\alpha_n$ as a function of the iteration index for one of our 100 inversions (panel (c)) and the RMSE and iteration number for ES-MDA-A2 in different configurations (panel (b)).

ES-MDA with eight iterations serves as the accuracy benchmark, with the BOD model clearly requiring many iterations to address nonlinearity.
ES-MDA-RLM and ES-MDA-RS both terminate earlier in just three iterations, at the expense of a higher RMSE; however, EKI-DMC performs comparable to the benchmark, with a slightly increased cost and reduced accuracy. In this test case, ES-MDA-A2 achieves the closest accuracy to the benchmark and requires on average only six iterations.

The configurability of ES-MDA-A2 is illustrated in Figure~\ref{fig:BODresults}(b), demonstrating that an increased $N_\text{max}$ improves accuracy and effectively bounds the iteration. When $N_\text{max}=3$ both ES-MDA-RLM and ES-MDA-RS are outperformed, and improvement continues until diminishing returns set in around $N_\text{max}=10$, where accuracy matches EKI-DMC and the benchmark.

The computational efficiency of ES-MDA-A2 arises from the adaptive inflation schedule in Figure~\ref{fig:BODresults}(c). ES-MDA-RLM and ES-MDA-RS decrease inflation quickly and terminate early, accepting a high RMSE. Conversely, EKI-DMC slowly decreases $\alpha_n$ each iteration, leading to better accuracy and a greater computational cost. ES-MDA-A2 starts with the smallest inflation because the proposal is based on RMSE and allows for increased $\alpha_n$ if the target RMSE is not yet reached. This strategy of aggressive early updates in combination with a catch-up mechanism proves highly effective here.

\begin{figure}[tb]
 \centering
 \includegraphics[width=1\linewidth]{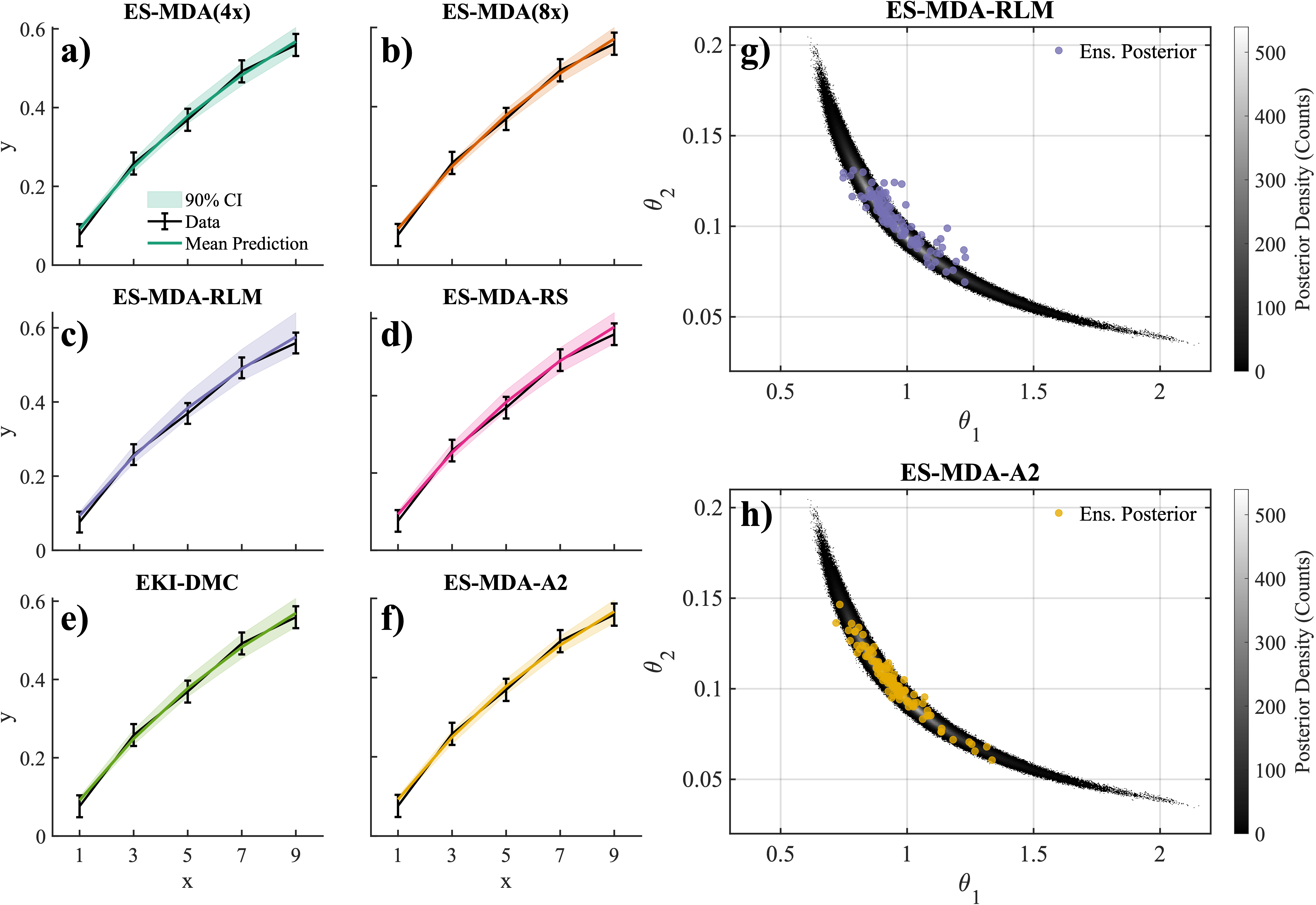}
 \caption{(a)-(f) Data fits of the final ensemble of ES-MDA methods.
 Solid lines correspond to the ensemble mean and shaded bands indicate $90\%$ confidence intervals (5th-95th percentiles).
 (a) ES-MDA with four iterations.
 (b) ES-MDA with eight iterations.
 (c) ES-MDA-RLM.
 (d) ES-MDA-RS.
 (e) EKI-DMC.
 (f) ES-MDA-A2.
 (g-h) Posterior distribution (heatmap), prior (grey dots) and posterior (orange dots) ensemble of (g) ES-MDA-RLM and (h) ES-MDA-A2.
 }
 \label{fig:BODdatafits}
\end{figure}
Figures~\ref{fig:BODdatafits}(a)-(f) show the final ensemble data fits and uncertainty for a single representative experiment. 
Panels (b), (e), and (f) are nearly identical, with ES-MDA(8x), EKI-DMC, and ES-MDA-A2 all fitting the data equally well, but ES-MDA-A2 requires the fewest iterations. ES-MDA-RLM's wider uncertainties in panel (c) reflect an early termination where the final ensemble may not be distributed according to the underlying Bayesian posterior distribution.
Figures~\ref{fig:BODdatafits}(g)-(h) show the posterior distribution as a heatmap computed via an affine invariant ensemble Markov chain Monte Carlo (MCMC) sampler using $10^6$ samples \cite{goodman2010ensemble,foreman2013emcee}.
We see that the posterior distribution over $\theta_1$ and $\theta_2$ is narrow and ``banana-shaped'' due to the nonlinearity of the BOD model.
ES-MDA-A2 captures this structure well (panel (h)), while several ensemble members of ES-MDA-RLM stray from the region of high posterior probability (panel (g)), consistent with the larger uncertainties.

\subsection{Electromagnetic inversions}
We now apply ES-MDA on two electromagnetic (EM) imaging problems, where we recover electrical resistivity as a function of depth.
Throughout, we estimate the logarithm of resistivity to avoid positivity constraints.
We consider two well-established data sets and corresponding forward models.
The first data set is from a Schlumberger DC resistivity sounding performed on the Australian shield \citep{constable1984deep}.
The second data set consists of seafloor magnetotelluric data, collected as part of a survey of
low-salinity groundwater in the continental shelf offshore New Jersey \citep{gustafson2019aquifer}.
Both data sets have been inverted many times and are often used to benchmark inversion methodologies in geophysics \citep[see, e.g.,][]{constable1987occam, RTOTKO}.

\subsubsection{Data, forward models and initial ensembles}
\begin{figure}[!htb]
 \centering
 \includegraphics[width=0.3\linewidth]{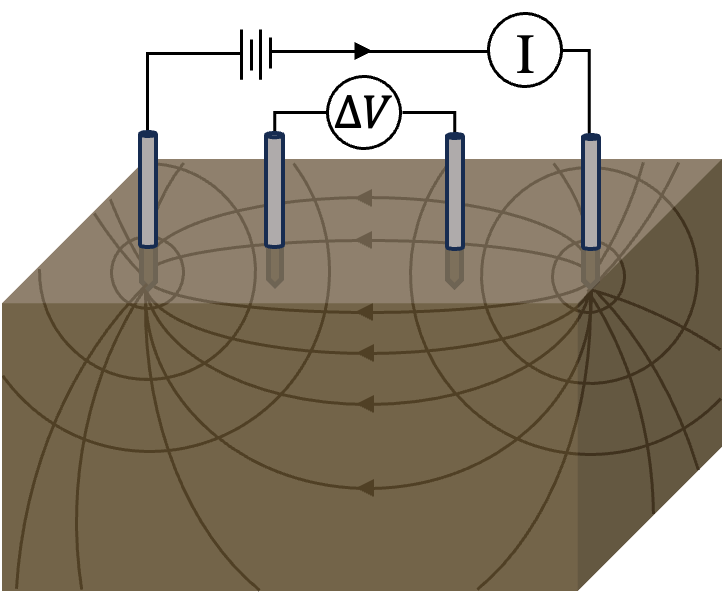}
 \includegraphics[width=0.25\linewidth]{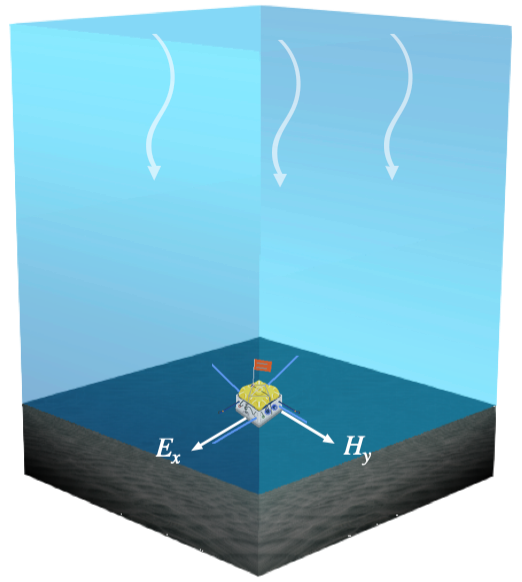}
 \caption{(a) Illustration of a Schlumberger sounding. 
 An outer pair of electrodes passes a DC current at exponentially increasing spacings.
 An inner pair of electrodes records measurements of the electric potential.
 (b) Illustration of a magnetotelluric (MT) receiver passively recording the magnetizing and electric fields on the seafloor.}
 \label{fig:EMIllustrations}
\end{figure}

The Schlumberger sounding is illustrated in Figure \ref{fig:EMIllustrations}(a): An outer pair of electrodes $AB$ passes a DC current $I$ at exponentially increasing spacings over 100 km, and an inner pair of electrodes $EF$ records measurements of the electric potential at each spacing.
The data observations are 29 apparent resistivities 
\begin{equation}
    \mathbf{d}_\text{obs} = (\rho_a^{(1)}, \rho_a^{(2)},\dots,\rho_a^{(29)})^\top,
\end{equation}
measured in Ohms ($\Omega$), where
\begin{equation}
    \rho_a^{(i)} = \frac{\pi(AB^{(i)}-EF)^2}{4I}\cdot\frac{\Delta V}{EF}.
\end{equation}
The forward model uses the filtering approach of \cite{ghosh1971application} to map resistivity on a logarithmically uniform grid with $M=60$ layers (down to a depth of $\sim$458 km) to apparent resistivity $\rho_a$.

The magnetotelluric (MT) method \citep{cagniard1953basic} estimates the frequency dependent complex impedance tensor
\begin{equation}\label{eq:imptensor}
    Z_{xy}(\omega) = \frac{H_x}{E_y},
\end{equation}
from orthogonal measurements of the horizontal magnetizing field $(H_x,H_y)$ and horizontal electric field $(E_x,E_y)$. 
Figure \ref{fig:EMIllustrations}(b) illustrates an MT receiver resting on the seafloor, passively acquiring magnetic and electric field measurements from the incident geomagnetic field and the associated induced electric field.
The frequency dependent complex impedance tensor $Z_{xy}$ defines a frequency dependent apparent resistivity $\rho_a$ and a corresponding phase $\phi$ via
\begin{equation}
    \rho_a = \frac{1}{\omega\mu_0}|Z_{xy}|^2,\quad \phi = \arctan\left(\frac{\text{Im}(Z_{xy})}{\text{Re}(Z_{xy})}\right),
\end{equation}
where $\mu_0$ is the magnetic permeability of free space. 
The field data are ten apparent resistivities and phase values at ten frequencies.
The forward model uses a standard recursion relationship \citep{ward1987electromagnetic} to map resistivity on a uniform grid with $M=60$ layers (each layer is 20m thick) to apparent resistivity and phase.

Initial ensembles of size $N_e=100$ are generated using multivariate normal distributions whose mean is the result of an Occam's inversion \citep{constable1987occam}
and whose covariance matrix $\mathbf{C}$ has elements
\begin{equation}
    C_{ij} = \exp{\left(-\left(\frac{r_{ij}}{\ell}\right)^2\right)},
\end{equation}
where $r_{ij}=r_{ji}=\|x_i-x_j\|$ is the distance between two grid points and $\ell$ is a length-scale.
We assign a different length scale to each ensemble member and draw length scales uniformly between 100m and 200m. 

\subsubsection{ES-MDA results: Accuracy vs. computational efficiency}
\begin{figure}[tb]
 \centering
 \includegraphics[width=\linewidth]{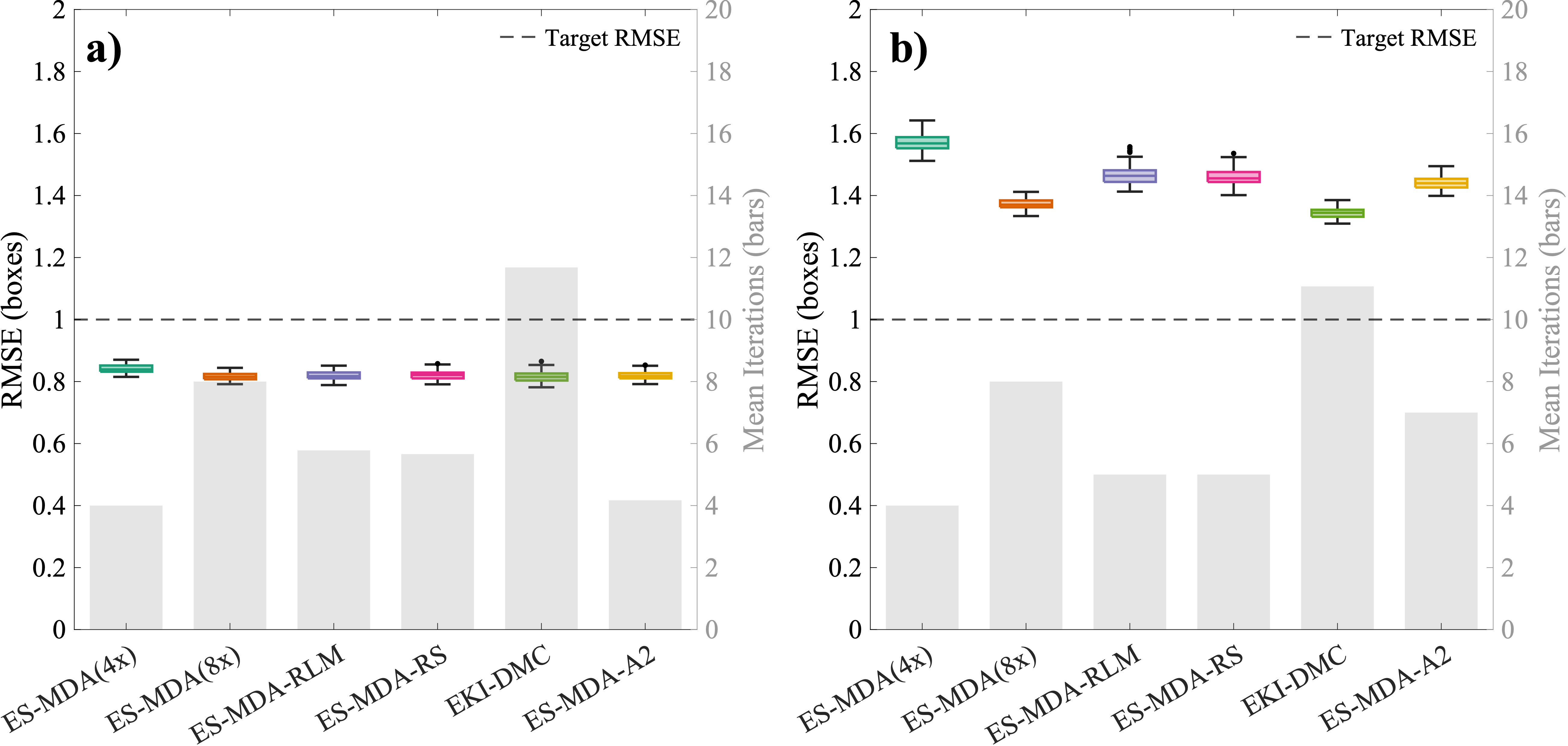}
 \caption{Summary of the numerical experiments with EM inversions.
 Shown are box-plots of RMSE (left y-axis) and bar-charts of the average number of iterations.
 For the box-plots: The box represents the interquartile range (IRQ), whiskers are calculated as $1.5\cdot \text{IQR}$, the median is shown as a solid colored line, and outliers shown as black dots
 (a) Results for DC resistivity inversions.
 (b) Results for MT inversions.}
 \label{fig:EMResults}
\end{figure}
Figure~\ref{fig:EMResults} summarizes cost and accuracy across 100 experiments for DC resistivity (panel (a)) and MT (panel (b)) inversions.

For DC resistivity, all methods easily achieve a final RMSE below one.
The adaptive schemes (ES-MDA-RS, ES-MDA-RLM, and ES-MDA-A2) match the eight-iteration benchmark at a lower cost and offer varied improvements over ES-MDA(4x). 
Here ES-MDA-A2 is the most efficient of the three and terminates by design upon reaching the targeted accuracy, requiring four iterations. 
EKI-DMC matches the eight-iteration benchmark in accuracy but incurs the highest computational cost, as its large inflation factors result in small model updates.

The MT results follow a pattern similar to that seen in the BOD test case.
ES-MDA-RS and ES-MDA-RML are cheapest but fall short of the eight-iterations benchmark in accuracy.
EKI-DMC again achieves high accuracy at the highest cost, for the same reason as above.
ES-MDA-A2 falls somewhere in-between this time, being slightly more expensive than ES-MDA-RS and ES-MDA-RLM, but more accurate, without quite reaching the benchmark accuracy.

\begin{figure}[!htb]
    \centering
    \includegraphics[width=0.7\linewidth]{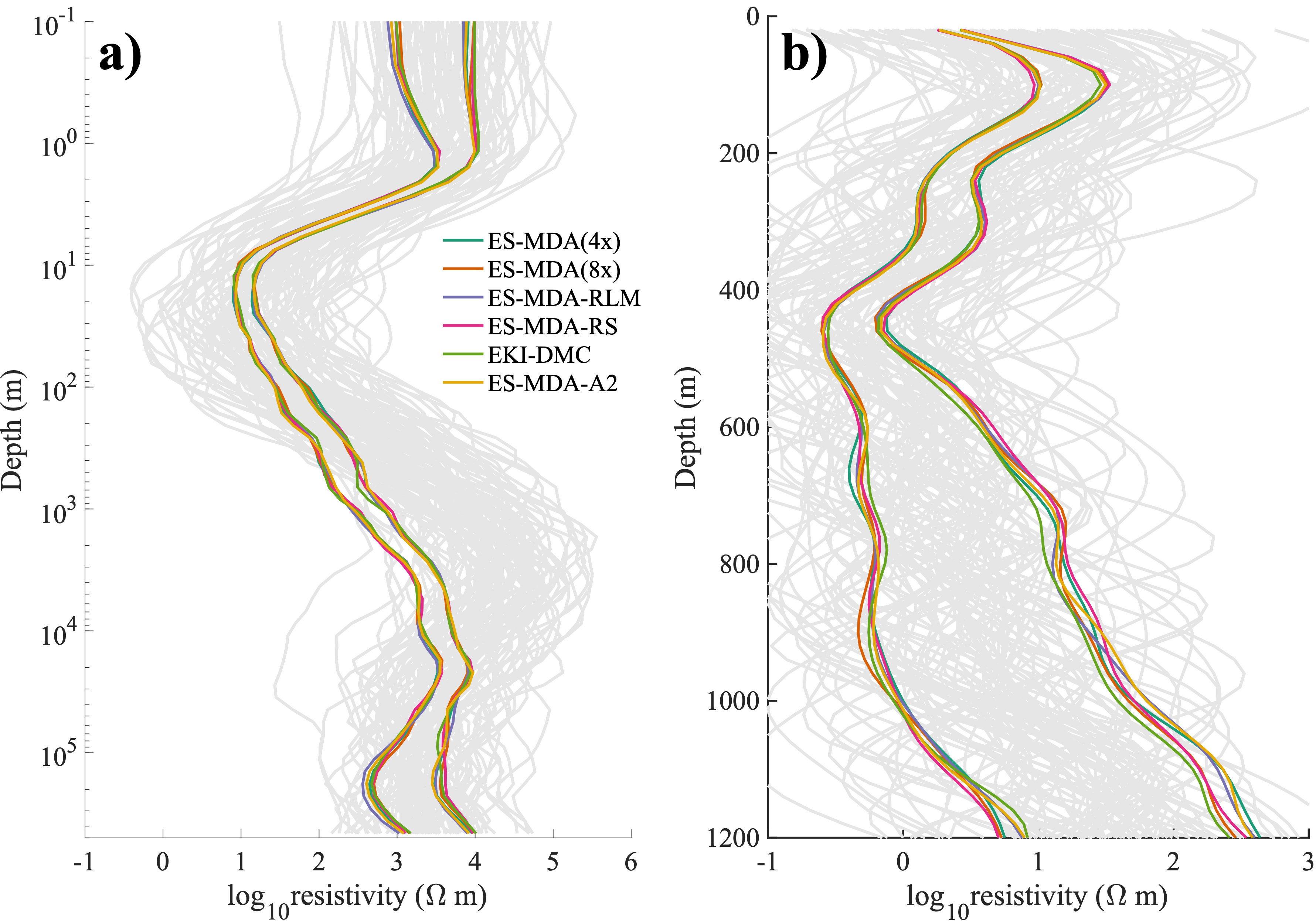}
    \caption{(a) DC resistivity inversion. (b) MT inversion. Colored lines define the posterior ensemble $90\%$ confidence interval for each method. Grey lines depict prior samples constituting the initial ensemble.}
    \label{fig:EMPosteriors}
\end{figure}

Figure~\ref{fig:EMPosteriors} illustrates that the various ES-MDA methods not only result in similar accuracies (final RMSE, as shown in Figure~\ref{fig:EMResults}), but also in similar posterior uncertainties. 
In the figure we show examples of prior and posterior ensembles ($90\%$ confidence intervals) for DC resistivity (panel~(a)) and MT inversions (panel~(b)).
Reassuringly, all methods lead to similar posterior uncertainties that are also comparable to uncertainties computed with other, computationally more costly methods \citep{RTOTKO}. Datafits for a single ensemble realization using ES-MDA-A2 can be found in Figure~S1 for the DCR case, and in Figure~S2 for the MT case.
Provided that all methods lead to comparable results, one should favor methods that are computationally less expensive.
The EM inversions we perform here suggest that adaptive ES-MDA methods have advantages over traditional ES-MDA with a fixed inflation schedule.
Among them, the way in which the inflation parameter is adapted during the iteration is what ultimately drives the accuracy-efficiency trade-off.

\subsection{2D reservoir injection}\label{sec:reservoir_sec}
\begin{figure}[tb]
 \centering
 \includegraphics[width=1\linewidth]{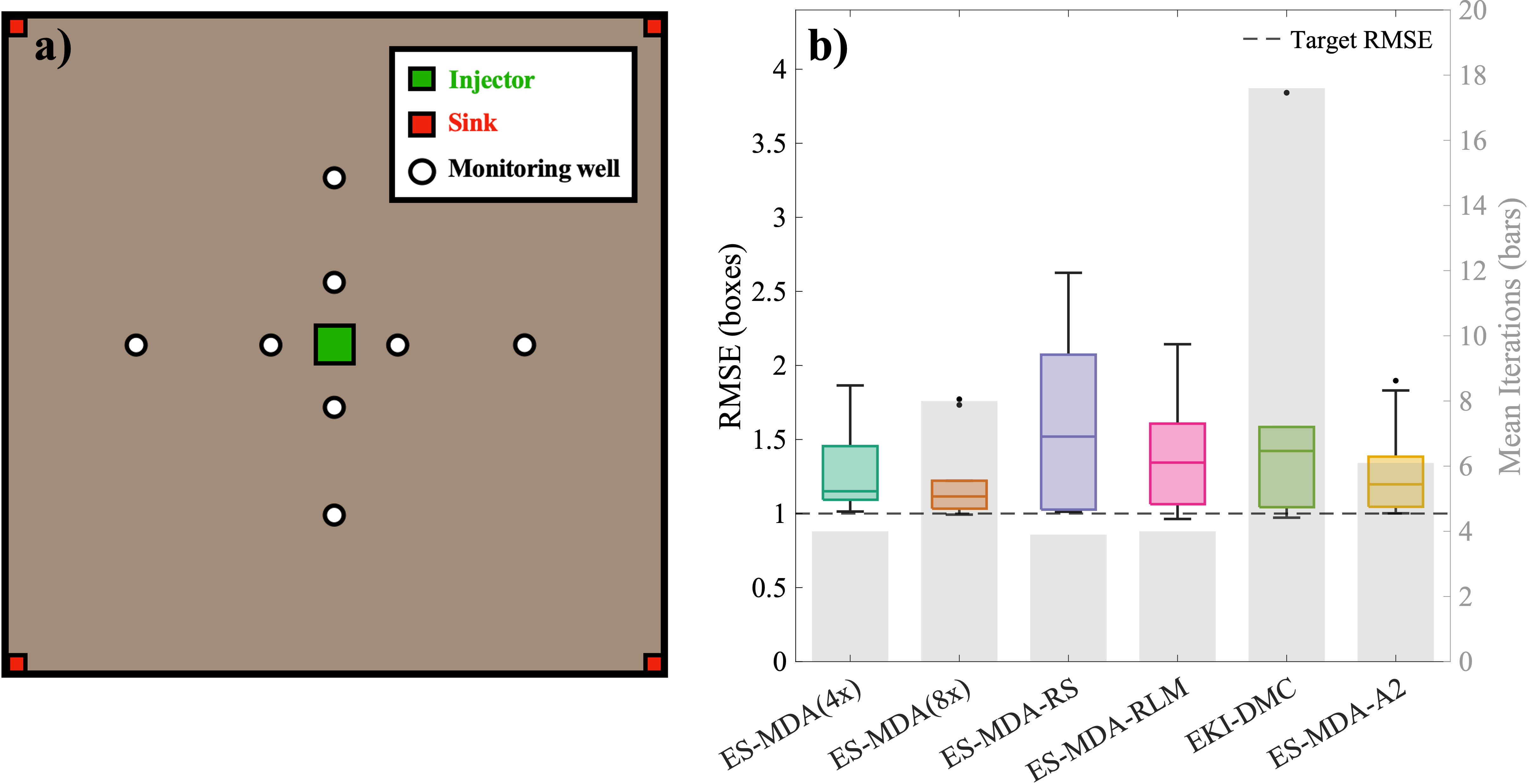}
 \caption{(a) 2D reservoir setup of the 9-spot problem hosting a central injector (green square) and eight monitoring wells (white) with data observations recorded at all well locations. (b) Colored box plots of RMSE overlain on a gray bar chart displaying the average number of iterations performed for each ES-MDA variant. We performed $N_\text{trials}=10$ with randomized initial ensembles of size $N_\text{e}=30$. The target RMSE one is plotted as a gray dashed line.
 }\label{fig:2Dreservoir_results}
\end{figure}

In the fourth test we consider the characterization of the subsurface permeability field of a synthetic 2D reservoir model. We study the 9-spot well configuration in Figure \ref{fig:2Dreservoir_results}(a), with a central injector and eight monitoring wells arranged in a cross pattern. Fluid flow is simulated using GEOS, an open-source, multi-physics software capable of modeling fluid flow and other processes in porous media \citep{settgast2024geos}. The behavior of a two-component (CO$_2$ and H$_2$O), two-phase (CO$_2$ and brine) reservoir is modeled at a depth of 2 km over a 20-year period. The 2D reservoir model extent is 10 km $\times$ 10 km and is discretized using a uniform grid of 100 m $\times$ 100 m grid cells. The inverse problem is to recover the $\mathbf{m}\in\mathbb{R}^{10,000}$ grid cell log-permeability parameters $\ln(k)$. The reference (``true'') reservoir permeability field is constructed from the convolution of a squared exponential kernel with a correlation length of 600 m and a Laplacian kernel with a correlation length of 1200 m, producing a spatially correlated Gaussian field with semi-linear features. To incorporate prior geological information, we condition this Gaussian process on known permeability values at a sparse set of locations defined on a coarse grid. The imposed conditioning enforces ensemble realizations that agree with the true permeability at these locations, while still allowing features to vary in between. History of the pressure $p$ and gas saturation $S_g$ are recorded biennially at the injection and monitoring wells, yielding $\mathbf{d}=\{(p_{t}^{(i)},S_{g,t}^{(i)})\in\mathbb{R}^2\mid i=1,\dots,9;t=1,\dots,10\}$. The observations are obtained by adding $5\%$ random noise to data generated with the reference permeability field. 

In Figure \ref{fig:2Dreservoir_results}(b) we report statistics from 10 experiments with an ensemble size of $N_e=30$ for this computationally expensive model. 
As in all previous numerical experiments, EKI-DMC incurs the largest computational cost due to the large inflation factors and small updates.
In this example, however, the large computational cost of EKI-DMC does not lead to a high accuracy -- the RMSE of EKI-DMC is higher than that of ES-MDA with eight iterations.
This finding suggests that many small updates are not always optimal for achieving a high accuracy. 

ES-MDA-RS and ES-MDA-RLM are computationally efficient (four iterations on average), but the accuracy is also low when compared to ES-MDA with eight iterations.
ES-MDA-A2 resolves the trade-off between accuracy and cost differently from the other adaptive methods. 
The cost is relatively low (about six iterations), and the accuracy is close to a more costly inversion with eight iterations of ES-MDA.

\begin{figure}[!htb]
 \centering
 \includegraphics[width=.75\linewidth]{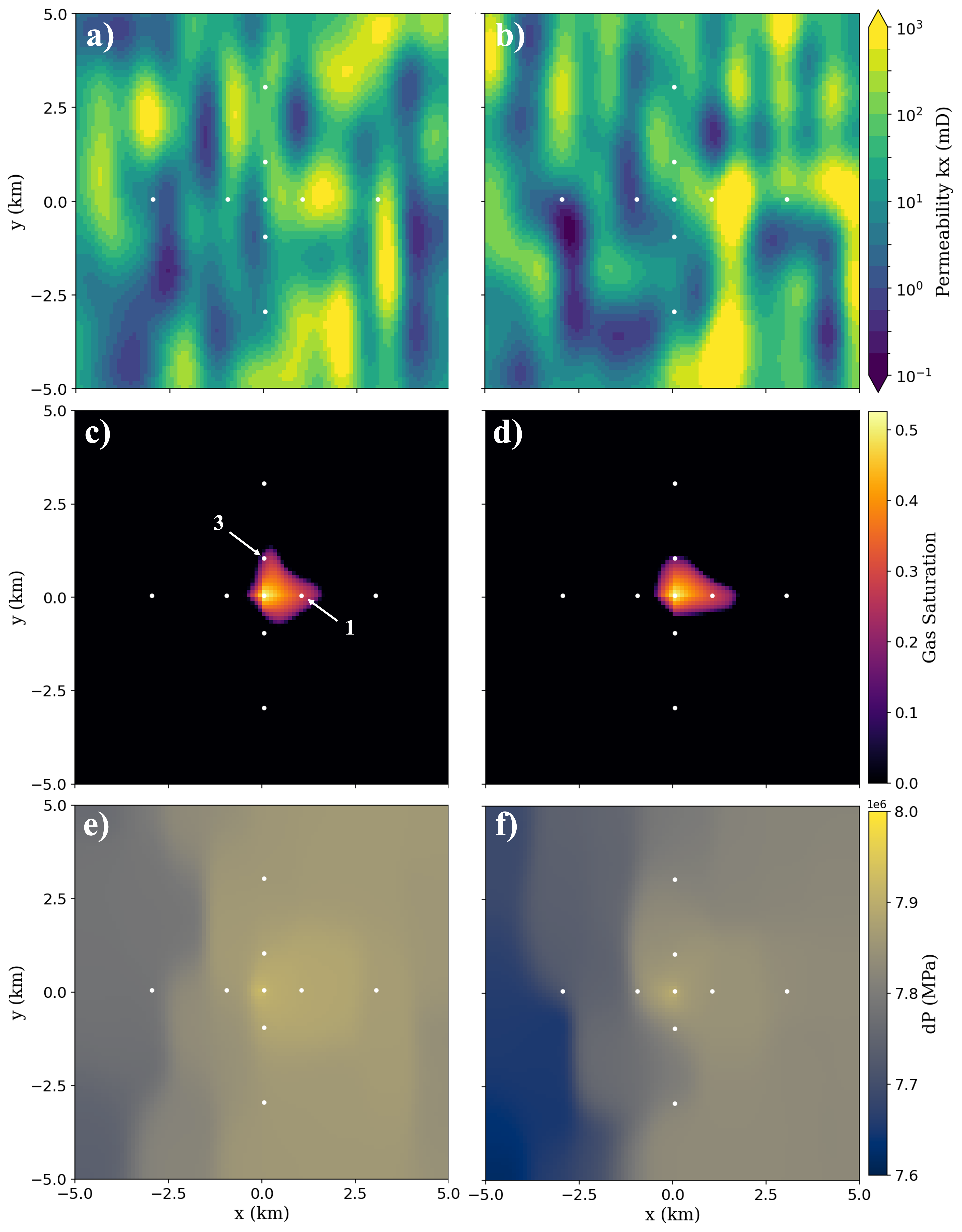}
 \caption{(a) ``True'' permeability used to generate synthetic data observations.
 (b) Recovered permeability field: posterior mean of ES-MDA-A2. 
 (c) ``True'' gas saturation at $t=20$ years.
 (d) Ensemble prediction of gas saturation at $t=20$ years (ES-MDA-A2). 
 (e) Change in ``true'' pressure  at $t=20$ years.
 (f) Change in pressure  at $t=20$ years (ES-MDA-A2).
 }
 \label{fig:2Dreservoir_gas}
\end{figure}

The true and recovered permeability fields using ES-MDA-A2 are shown in Figures \ref{fig:2Dreservoir_gas}(a)-(b) for a single experiment.
The inversion captures the major structures of the true permeability field, although discrepancies remain in the far-field regions, which are poorly constrained by the small number of data observations we use for the inversion. 
True and recovered gas plume migrations at $t=20$ years are shown in Figures \ref{fig:2Dreservoir_gas}(c)-(d) and we note that ES-MDA-A2 accurately tracks the northeastward plume movement. Additionally, the true and recovered changes in the pressure field at $t=20$ years are shown in Figures\ref{fig:2Dreservoir_gas}(e)-(f) where the general east to west pressure gradient is captured. The pressure field is effectively resolved in the constrained model region, with small deviations from the reference in the far-field. 

 The associated data fits using ES-MDA-A2 are shown in the Supplementary Figures~S3 (gas saturation) and S4 (pressure).
 We note that ES-MDA-A2 matches the change in pressure observed at all wells (despite getting the permeability field wrong in the poorly-constrained far-field).
 Moreover, ES-MDA-A2 captures the gas saturation trajectories at the injection well and monitoring wells 3 and 5 correctly. Importantly, no spurious gas saturation is predicted at the wells which have never encountered the plume, although the prior ensemble includes such scenarios.

\section{Conclusions}
\label{sec:conclusions}
In this work, we showed how ES-MDA entails an inherent tradeoff of accuracy and computational efficiency for large-scale inverse problems. 
Defining the number and size of updates a priori, as is often done in practice, can require several trials before a satisfactory data fit is obtained and is computationally wasteful, especially when history matching field-scale models. 
Adaptive ES-MDA variants address this issue by selecting the number and size of the updates during the iteration.

We describe a unifying framework for ES-MDA inflation scheduling and identify three steps that differentiate ES-MDA methods: inflation proposal, pre-analysis inflation revision, and post-analysis inflation revision. 
Small inflation parameters yield larger updates and greater effiency, while large inflation parameters yield small updates and greater accuracy.
Most existing adaptive ES-MDA variants favor either efficiency \emph{or} accuracy. 
Methodological differences can primarily be attributed to pre- and post-analysis revision, where conditional checks enforce a change in the inflation proposal. 
In effect, the inflation proposal sets the pace for convergence and it is the inflation revisions that modulate the pace. 

Using insights from the three-step framework, we introduce a new adaptive method, ``accuracy-aware'' ES-MDA (ES-MDA-A2, Section~\ref{sec:ESMDAA2}), which performs large early updates but allows additional smaller updates when a target accuracy is not reached, using only two interpretable inputs: a target RMSE and a maximum number of iterations. In our experiments, ES-MDA-A2 resolves the accuracy-efficiency trade-off differently from existing methods: it matches the accuracy of the most accurate adaptive method at reduced computational cost, and its two inputs allow it to be configured to behave like existing methods or to yield ``accurate enough'' results at low cost.

The systematic numerical experiments that support our findings range from toy models, to 1D inversions of electromagnetic (EM) data, to 2D inversions with a multi-physics reservoir simulator. In the toy model, we demonstrate that ES-MDA-A2 can be configured to match or exceed the accuracy of other adaptive methods through the choice of the maximum number of iterations. 
In the EM inversions, ES-MDA-A2 achieved a similar accuracy to other methods at a lower cost, or it landed in between the state-of-the-art in terms of accuracy and cost. 
Lastly, in a 2D reservoir test case, ES-MDA-A2 outperforms other adaptive ES-MDA methods in terms of both accuracy and computational cost. Interestingly, in the reservoir case EKI-DMC was both less accurate than ES-MDA-A2 and the most expensive of the methods, suggesting that performing many small updates is not always optimal. 
This is understandable from a sampling error perspective, where each update accumulates error derived from the small ensemble size until the updated is dominated by noise, an idea explored in \cite{evensen2018analysis}. In addition, it is clear that inflation revision both before and after the analysis step is useful in dictating ES-MDA convergence. 
From our experiments, we find that sufficient iteration to address nonlinearity is critical and that there is perhaps more flexibility in inflation schedule evolution than previously thought. 

\bmhead{Supplementary information}

Four supplementary figures are provided, showing datafits for a single ensemble trial of ES-MDA-A2 of the DCR, MT, and 2D reservoir test cases. MATLAB code implementing the ES-MDA algorithm variants explored in this work, along with code to reproduce the BOD model results, are included as supplementary information.

\bmhead{Acknowledgments}

The authors thank Lawrence Livermore National Laboratory for providing high performance computing resources and software assistance, and gratefully acknowledge support from the U.S. Department of Energy, Hydrocarbons and Geothermal Energy Office through Field Work Proposal FEW0301 (RamonCO), and from Lawrence Livermore National Laboratory under Contract DE-AC52-07NA27344.
MM  is supported by the U.S. Office of Naval Research Grant N000142512298.

\section*{Declarations}

\bmhead{Funding}
This work was supported by the U.S. Department of Energy, Hydrocarbons and Geothermal Energy Office through Field Work Proposal FEW0301 (RamonCO).  Portions of this work were performed under the auspices of the U.S. Department of Energy by Lawrence Livermore National Laboratory under Contract DE-AC52-07NA27344.

\bmhead{Competing interests}
The authors have no competing interests to declare that are relevant to the content of this article.

\bmhead{Ethics approval and consent to participate}
Not applicable.

\bmhead{Consent for publication}
Not applicable.

\bmhead{Data availability}
The BOD model and data used in this study are described in~\cite{bardsley2014randomize}. The magnetotelluric data used in this study are investigated in~\cite{gustafson2019aquifer} and publicly available from the Lamont-Doherty Earth Observatory EM Lab freshwater project repository at \url{https://emlab.ldeo.columbia.edu/index.php/projects/freshwater/}. The DCR Schlumberger sounding data are reported in~\cite{constable1984deep}. Data for the reservoir example are available from the corresponding author upon reasonable request.

\bmhead{Materials availability}
Not applicable.

\bmhead{Code availability}
The MATLAB code implementing the ES-MDA algorithm variants discussed in this work, along with code to reproduce the BOD model results, will be provided on Zenodo.
The multi-physics reservoir simulator GEOS is open-source~\cite{settgast2024geos}; reproducing the reservoir example requires a user-supplied wrapper to interface with GEOS, which is not included here.

\bmhead{Author contribution}
K.I. and M.M. wrote the main manuscript text and conducted the numerical experiments therein. K.I., M.M., and J.A.W. conceptualized the contents of the manuscript. K.I. prepared all figures within the text. K.I., M.M., C.S.S, and C.M. wrote the code and provided software assistance necessary for the numerical experiments. M.M., R.M., and J.A.W. provided supervision and funding acquisition. All authors reviewed the manuscript.

\bibliography{sn-bibliography}

\end{document}


\maketitle

\noindent \textit{Journal: Computational Geosciences}\\
\noindent \textit{Corresponding authors: Kyle Ivey (krivey@ucsd.edu), Matthias Morzfeld (mmorzfeld@ucsd.edu)}\\
\noindent \textit{Affiliation: Scripps Institution of Oceanography, University of California, San Diego, 9500 Gilman Drive, La Jolla, 92093, CA, USA}

\vspace{1em}

\begin{figure}[!htb]
 \centering
 \includegraphics[width=.9\linewidth]{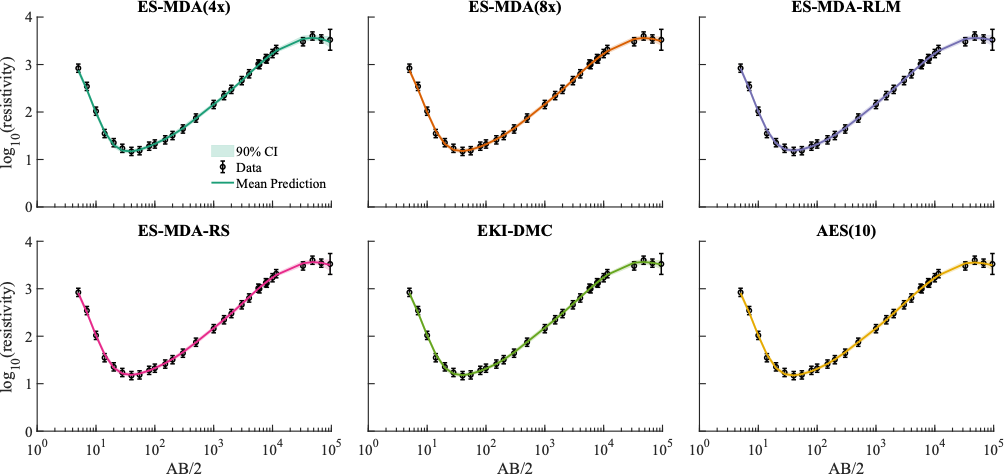}
 \caption{DCR data fits of the final ensemble for each ES-MDA method.}
 \label{fig:S1_DCR}
\end{figure}

\begin{figure}[!htb]
 \centering
 \includegraphics[width=.9\linewidth]{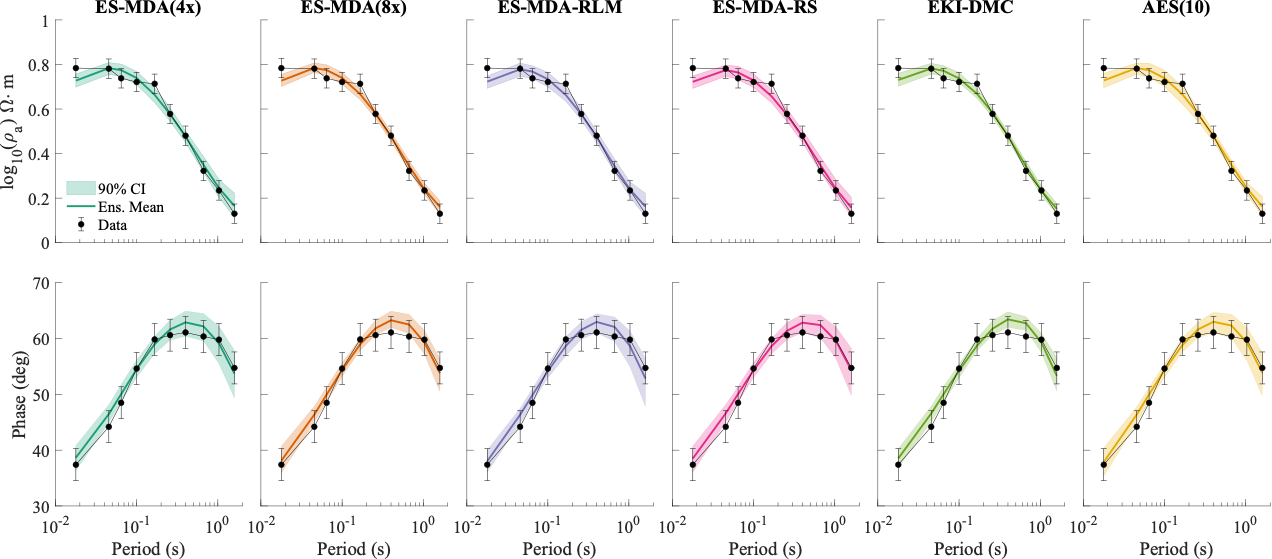}
 \caption{MT data fits of the final ensemble for each ES-MDA method.}
 \label{fig:S2_MT}
\end{figure}

\begin{figure}[!htb]
 \centering
 \includegraphics[width=.9\linewidth]{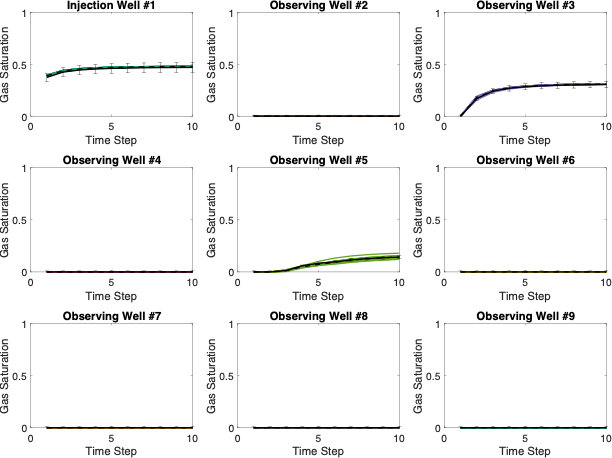}
 \caption{2D reservoir data fits of gas saturation at each well with the final ensemble for each ES-MDA method.}
 \label{fig:S3_gas}
\end{figure}

\begin{figure}[!htb]
 \centering
 \includegraphics[width=.9\linewidth]{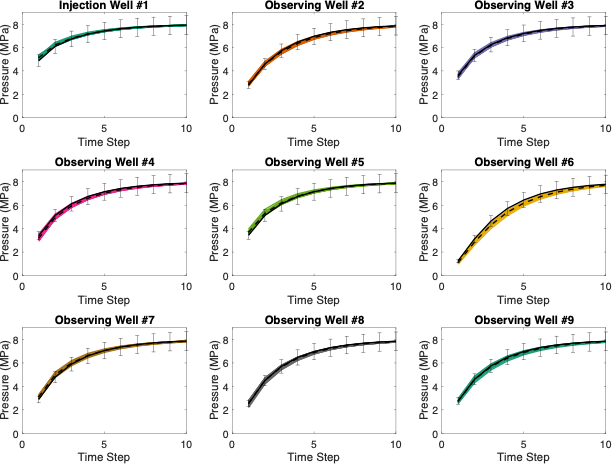}
 \caption{2D reservoir data fits of pressure at each well with the final ensemble for each ES-MDA method.}
 \label{fig:S4_pressure}
\end{figure}